# Unusual interlayer quantum transport behavior caused by the zeroth Landau level in YbMnBi$_2$

J.Y. Liu[1], J. Hu[1*], D. Graf[2], T. Zou[3], M. Zhu[3], Y. Shi[4], S. Che[4], S.M.A. Radmanesh[5], C.N. Lau[4], L. Spinu[5], H.B. Cao[6], X. Ke[3] and Z.Q. Mao[1*]

[1]Department of Physics and Engineering Physics, Tulane University, New Orleans, LA 70118, USA

[2]National High Magnetic Field Lab, Tallahassee, FL 32310, USA

[3]Department of Physics and Astronomy, Michigan State University, East Lansing, MI 48824, USA

[4]Department of Physics, University of California, Riverside, CA 92521, USA

[5] Department of Physics and Advanced Materials Research Institute, University of New Orleans, New Orleans, LA 70148

[6]Quantum Condensed Matter Division, Oak Ridge National Laboratory, TN 37830, USA.

## Abstract

**Relativistic fermions in topological quantum materials are characterized by linear energy-momentum dispersion near band crossing points. Under magnetic field, relativistic fermions acquire Berry phase of $\pi$ in cyclotron motion, leading to a zeroth Landau level (LL) at the crossing point [1]. Such field-independent zeroth LL, which distinguishes relativistic fermions from conventional electron systems, is hardly probed in transport measurements since the Fermi energy ($E_F$) is usually not right at the band crossing points in most topological materials. Here we report the observation of exotic quantum transport behavior resulting from the zeroth LL in a multiband topological semimetal YbMnBi$_2$ which possesses linear band crossings**

**both at and away from the Fermi level (FL) [2]. We show that the Dirac bands with the crossing points being above or below the FL leads to Shubnikov de-Haas oscillations in the in-plane magnetoresistance, whereas the Dirac bands with the crossing points being at the FL results in unusual angular dependences of the out-of-plane magnetoresistance and in-plane Hall resistivity due to the dependence of the zeroth LL's degeneracy on field orientation. Our results shed light on the transport mechanism of the zeroth LL's relativistic fermions in layered materials.**

[*]jhu@tulane.edu; zmao@tulane.edu

In conventional metals, the energy of the quantized landau level (LL) increases linearly with increasing magnetic field. In contrast, in topological materials (*e.g.*, Dirac and Weyl semimetals), the quantized energy states are given by $\varepsilon_n = \pm\upsilon_F\sqrt{2e\hbar B|n|}$ ($n$=0, ±1, ±2…), with $n$ = 0 corresponding to the zero energy LL ($n$=0) [1]. Such a zeroth LL, which is absent in non-relativistic electron systems, manifests itself in observable effects such as the phase shift (*i.e.* Berry phase) in quantum oscillations [3,4]. Quantum oscillations are caused by the oscillatory density of states at the Fermi energy (DOS($E_F$)) due to the LLs successively passing through the Fermi energy $E_F$. In a topological material, the zeroth LL is pinned to the Dirac crossing point regardless of magnetic field strength. When the linear Dirac bands cross at $E_F$, the increase of degeneracy of zeroth LL upon increasing field would lead to a monotonic increase of DOS($E_F$), which causes monotonically enhanced conductivity rather than quantum oscillations. A direct probe of such relativistic fermion transport is not feasible for recently-discovered, three dimensional (3D) Dirac/Weyl semimetals such as $Cd_3As_2$ [5-8], $Na_3Bi$ [9,10] and TaAs-type monopnictides [11-17], since the Dirac/Weyl nodes of these materials are not right at $E_F$.

The recently reported topological semimetal $YbMnBi_2$ [2], however, provides an excellent opportunity to study transport behavior of the zeroth LL's relativistic fermions. $YbMnBi_2$ features a layered structure, with the Bi square net layers harboring relativistic fermions. Angle-resolved photoemission spectroscopy (ARPES) studies have revealed that its Fermi surface consists of hole and electron pockets; both pockets are comprised of Dirac bands, with the band crossing points being below/above the FL. At the connection points of electron and hole pockets, cone-like dispersions with the node being at $E_F$ appear, which is believed to result from a time-reversal symmetry breaking type-II Weyl state [2]. Although evidence for time-reversal symmetry breaking remains to be elucidated [18], the linear Dirac band crossings at $E_F$ enable the study of the zeroth LL through transport measurements. In this letter, we show the zeroth LLs of the Weyl cones in $YbMnBi_2$ cause exotic quantum phenomena in both the out-of-plane magnetotransport and in-plane Hall effect, which has never been observed in other 3D topological semimetals.

The $YbMnBi_2$ single crystals were synthesized using a flux method (see Methods). We have performed neutron scattering experiments on $YbMnBi_2$ single crystals, which not only confirmed its tetragonal lattice structure (see Supplementary Table 1 for detailed structural parameters), but also revealed a C-type AFM state below $T_N = 298$ K, with the ordered moment of 3.789(3) $\mu_B$/Mn (Fig. 1a and 1b), in agreement with the magnetic structure reported previously by Wang et al. [18]. The Yb spins do not order even down to 4K. Although very weak ferromagnetism was seen in the magnetization measurements of YbMnBi$_2$ (see in Supplementary Fig. 2), it could not be resolved in neutron scattering experiments within the instrumental resolution.

Signatures of relativistic fermions in YbMnBi$_2$ have been observed in transport measurements. As shown in Fig. 1c, both the in-plane ($\rho_{xx}$) and out-of-plane ($\rho_{zz}$) resistivity display metallic temperature dependence with the anisotropic ratio $\rho_{zz}/\rho_{xx}$ reaching 36 at $T$ = 2 K, suggesting a moderately anisotropic electronic structure. Upon applying magnetic field along the out-of-plane direction (c-axis of YbMnBi$_2$), YbMnBi$_2$ displays large magnetoresistance (MR) for $\rho_{xx}$, ~$1.5\times10^4$ % at 45T and 2K (Fig. 1d). At low fields, MR displays a quadratic field dependence, which is generally expected for the classical Lorentz effect. Nevertheless, a linear dependence develops above a few tesla at high temperatures and persists up to 125K (inset of Fig. 1d), which is widely seen in Dirac/Weyl semimetals [19-22].

At low temperatures, SdH oscillations become visible above $B$ =15T (Fig. 1d). Two oscillation frequencies, $F_\alpha$=115T and $F_\beta$ = 162T, can be extracted from the fast Fourier transform (FFT) for the oscillatory component $\Delta\rho_{xx}$ (see the inset to Fig. 1e), consistent with the previous report [18]. The effective cyclotron masses $m^*$ associated with these two frequencies are ~ $0.24m_0$ ($m_0$, the free electron mass), which are obtained by fitting the temperature dependence of the FFT amplitude to the thermal damping factor of the Lifshitz-Kosevich (LK) formula [23,24], *i.e.*, $\frac{2\pi^2 k_B T m^* / \hbar eB}{\sinh(2\pi^2 k_B T m^* / \hbar eB)}$, as shown in Fig. 1e.

As noted above, the Berry phase of $\pi$ accumulated in cyclotron motion is the fundamental topological property of relativistic fermions. However, for YbMnBi$_2$, the Berry phase cannot be precisely determined using the commonly used LL fan diagram due to the existence of multiple frequencies in its SdH oscillations, but is accessible through the direct fit of the oscillation pattern to the multiband LK formula [25] (see Methods). The Berry phases of YbMnBi$_2$ obtained in our two-band LK fit (Fig. 1f) is $0.8\pi$ for $F_\alpha$ bands and $-0.6\pi$ for $F_\beta$ bands. This result is based on the assumption that both $F_\alpha$ and $F_\beta$ bands are 2D. Given that the electronic band structure of YbMnBi$_2$ is just moderately anisotropic as noted above, its dimensionality of band structure may be between 2D and 3D. In the 3D case, the Berry phase would be $0.8\pi \pm 0.25\pi$ for $F_\alpha$ bands and $-0.6\pi \pm 0.25\pi$ for $F_\beta$ bands. In either 2D or 3D case, the fitted Berry phases are clearly non-trivial. The small deviation from the fit near $1/B \sim 0.03$ in Fig. 1f is possibly due to the fact that we did not consider higher harmonic components and/or Zeeman splitting in the fit.

The observed SdH oscillations should arise from the Dirac cones with the nodes being away from $E_F$. However, the Weyl cones with the nodes being at the FL at the connection points of electron and hole pockets would not contribute to the SdH oscillations. Since the zeroth LLs of such Weyl cones are pinned to the FL, its increased degeneracy upon increasing magnetic field would cause DOS($E_F$) to increase monotonically as noted above. This effect, though hardly resulting in noticeable features in the in-plane magnetoresistance, yields peculiar signatures in the field orientation dependence of interlayer magnetoresistance and in-plane Hall resistivity, as will be discussed below.

As noted above, the relativistic fermions in YbMnBi$_2$ are generated by 2D Bi planes which are separated by Yb-MnBi$_4$-Yb layers (Fig. 2a). The interlayer electron transport is expected to be dominated by a coherent momentum relaxation mechanism if the Yb-MnBi$_4$-Yb slab layer is conducting or by tunneling processes if the Yb-MnBi$_4$-Yb layer is insulating or less conducting. Our observation of the moderate resistivity anisotropy ($\rho_{zz}/\rho_{xx} \sim 36$) implies that the Yb-MnBi$_4$-Yb layers are less conducting. To capture the essential physics, here we first assume a 2D limit by considering insulating Yb-MnBi4-Yb slabs. Under this assumption, the interlayer transport is supposed to be dominated by tunneling processes, as depicted in Fig. 2a. As indicated above, ARPES measurements on YbMnBi$_2$ have shown its Weyl nodes are right at the FL. Therefore, the LL quantization of the Weyl cones under out-of-plane magnetic fields leads the zeroth LLs to be pinned to the FL. In this case, the tunneling current of the zeroth LL's relativistic fermions should enhance with increasing the magnetic field, owing to the increased LL degeneracy. Given that the electron cyclotron motion is confined within the plane in such a 2D limit, rotating field away from the out-of-plane direction would suppress LL quantization, hence reducing the tunneling conductivity. Such a phenomenon has been demonstrated in the pressurized layered organic conductor α-(BEDT-TTF)$_2$I$_3$, which has a 2D Dirac cone with the node being exactly at $E_F$ in each BEDT-TTF molecular layer [26,27].

While the electronic structure in YbMnBi$_2$ is not exactly 2D, its anisotropic band structure enables the observation of signatures of zeroth LL relativistic fermion tunneling. As shown in Fig. 2b, the interlayer MR becomes the most remarkable when $B \perp I$. With reducing the polar angle (see the inset of Fig. 2b for the experimental setup), MR is quickly suppressed and shows sub-linear field dependence. Such an unusual evolution of MR with $\theta$ cannot be understood in light of the MR induced by the classical orbital effect or by other quantum effects such as weak anti-localization as discussed in Supplementary Note 1. However, it is consistent with the expectation that the interlayer tunneling of the zeroth LL's relativistic fermions enhances interlayer conductivity as the field is rotated toward the out-of-plane direction. We will show below that the sublinear field dependence of MR observed for $\theta < 90°$ can indeed be quantitatively described by the tunneling model of the zeroth LL's relativistic fermions.

The essential evidence for the interlayer tunneling of the zeroth LL's relativistic fermion were found in the measurements of the angular-dependence of interlayer magnetoresistance (AMR) and in-plane Hall resistivity. We present the AMR data under various magnetic fields at 2K in Fig. 2c where AMR is defined as $\Delta\rho_{zz}(\theta,B)/\rho_{zz,\min}$ = $[\rho_{zz}(\theta,B)-\rho_{zz}(0,B)]/\rho_{zz}(0,B)$. Noticeably, the AMR at 31T displays a very sharp peak at $\theta = 90°$ ($B \perp I$) and is nearly angle-independent below $\theta = 60°$. With decreasing the field, the peak becomes suppressed and broadened; significant suppression and broadening are observed below 9T. Surprisingly, when $B \leq 1T$, MR ($\theta$) evolves into $\sin^2\theta$ dependence as shown by the solid fitted curves, in stark contrast with MR($\theta$) at high fields. The $\sin^2\theta$ dependence of MR($\theta$) is generally expected for the classical orbital effect for which

MR($\theta$) $\propto B_{xy}^2 = B^2\sin^2\theta$. The strong deviation of MR($\theta$) from the $\sin^2\theta$ dependence in the high field range implies that the interlayer transport mechanism in the high field range is distinct from that in the low field range.

Although there are several known mechanisms which can result in unusual anisotropy in MR such as spin scattering, quantum interference effect and Fermi surface anisotropy, none of them can offer a reasonable interpretation for our observed AMR anisotropy in the high field range as discussed in Supplementary Note 1. Instead, the model of the interlayer tunneling of the zeroth LL's relativistic fermions can quantitatively describe our AMR data in Fig. 2b and 2c. As shown in Ref. [26], the tunneling conductance $\sigma_t^{LL0}$ due to the zeroth LL in a multilayer relativistic fermion system can be described by:

$$\sigma_t^{LL0} = A \cdot |B\cos\theta| \exp[-\frac{1}{2}\frac{ed^2(B\sin\theta)^2}{\hbar|B\cos\theta|}] \qquad (1)$$

where $A$ is a field independent parameter, $d$ = 1.0824 nm is the interlayer spacing of the neighboring Bi layers. In YbMnBi$_2$, the moderately anisotropic transport properties, as well as the existence of Dirac bands with the nodes being away from the FL as stated above, imply an additional coherent momentum relaxation channel $\sigma_c$ in the interlayer transport, which yields a classical quadratic term in low field MR. Thus the conductance of this channel under a low magnetic field can be expressed as $\sigma_c = \sigma_0/(1+k_1 \cdot B_{xy}^2)$, where $\sigma_0$ is the Drude conductivity and $k_1$ is a per-factor. At a sufficiently high field where all Dirac fermions are condensed to the lowest $n$ = 0 LL (namely reaching the quantum limit), a linear MR is expected and $\sigma_c$ can be rewritten as $\sigma_0/(1+k_2 \cdot |B_{xy}|)$ [28].

For YbMnBi$_2$, due to the existence of Weyl cones with its nodes being at the $E_F$, the quantum limit is easily achieved when the spacing between the zeroth and first LL is greater than the breadth of LL. In YbMnBi$_2$, we have observed linear MR in both $\rho_{xx}$ and $\rho_{zz}$ develops above a few tesla for $B \perp I$. Taking the linear and quadratic field dependence of $1/\sigma_c$ for high ($B>3T$) and low field ($B<3T$) respectively, our AMR data can be well fitted using $\rho_{zz}(\theta)=1/(\sigma_t^{LL0}+\sigma_c)$, as shown in Fig. 2c where the black and red solid lines represent the fitted curves for high and low fields, respectively. At lower fields where the tunneling is weak due to the suppressed LL quantization, the interlayer transport is dominated by the conventional conduction channel where the MR and AMR follow $B^2$ and $\sin^2\theta$ dependence respectively, which is expected for the classical orbital effect. This is exactly what we have observed for $B \leq 1T$ as described above. Additionally, the sub-linear behavior of the field dependence of MR in Fig. 2b can also be quantitatively fitted to $\rho_{zz}(\theta)=1/(\sigma_t^{LL0}+\sigma_c)$ as shown by the black solid fitted curves, which further validates our model.

In addition to the tunneling process discussed above, we find the conduction through the momentum relaxation channel alone can also account for the AMR data. The conductance due to the zeroth LL Dirac fermions through the relaxation channel can be written as $\sigma_c^{LL0}=\sigma_0^{LL0}/(1+k_1 \cdot B_{xy}^2)$ for low fields (<3T) and $\sigma_c^{LL0}=\sigma_0^{LL0}/(1+k_2 \cdot |B_{xy}|)$ for high fields (>3T). When the zeroth LL is locked to the FL, which is the case for the

Weyl cone in YbMnBi$_2$, $\sigma_0^{LL0} \propto DOS(E_F) \propto B_z$ given that the LL degeneracy is proportional to $B_z$. Replacing the tunneling term $\sigma_t^{LL0}$ by $\sigma_0^{LL0}$ in eq. (1), we obtain $\rho_{zz}(\theta) = 1/(\sigma_c^{LL0} + \sigma_c)$, which can also reproduce the AMR data at 31T, as shown in the inset to Fig. 2c. But the fit is less satisfied as compared to the fit by the tunneling model. It should be emphasized that in both models the unusual AMR is determined by the dependence of the zeroth LL's degeneracy on the magnetic field and its orientation.

The interlayer quantum transport of the zeroth LL's relativistic fermions discussed above is further corroborated by our measurements of the dependence of in-plane Hall resistivity on field orientation. We note such an experimental approach was used to demonstrate the interlayer tunneling of the zeroth LL' relativistic fermions in the pressurized layered organic conductor α-(BEDT-TTF)$_2$I$_3$ [29,30]. Fig. 3a shows our experimental setup; the in-plane transverse (x-axis) Hall voltage is measured with the out-of-plane (z-axis) current, and the magnetic field of fixed strength is rotated within the yz-plane. In a simple metal, Hall resistivity for such an experiment set up is given by $B_y/ne$, where $B_y = B\sin\theta$ is the field component perpendicular to current, and $n$ is the carrier density. This leads the Hall resistance $R_{zx}$ to follow a $\sin\theta$ dependence with the rotation of the field, which is indeed observed in YbMnBi$_2$ for weak fields ($B<1$T), as shown in Fig. 3b. However, $R_{zx}(\theta)$ starts to deviate from the $\sin\theta$ dependence for $B>2$T and such a deviation becomes significant for $B > 6$T, as shown in Fig. 3c, which can be understood in terms of the formation of zeroth LL. As discussed above, when the energy spacing between the zeroth and 1st LL is greater than the LL's breadth, the DOS($E_F$) contributed by the bands with linear crossing at $E_F$ should monotonically increase upon increasing field and is proportional to the out-of-plane field component $B\cos\theta$. Therefore, a $\tan\theta$ dependence is expected for $R_{zx}(\theta)$ since $\rho_H \propto B_y/ne \propto B\sin\theta/B\cos\theta = \tan\theta$. Indeed, we observed such a dependence, as shown in Fig. 3d where $R_{zx}(\theta)$ is plotted against $\tan\theta$. It is interesting to note that $R_{zx}(\theta)$ measured at different fields collapse into a single line (*i.e.* the black dashed line in Fig. 3d) in a lower angle region, which is not surprising, since $\rho_{zx} \propto \tan\theta$ is field independent. At large angles, Landau quantization is suppressed due to reduced $B_z$, causing the deviation from the $\tan\theta$ asymptote. The deviation angle is larger for higher fields, since the threshold field, $B_{c,z}=B\cos\theta_c$, for the distinguishable zeroth LL can be satisfied at higher angles.

## Methods

### Single Crystal Preparation

The YbMnBi$_2$ single crystals were synthesized using a self-flux method with the stoichiometric mixture of Yb, Mn and Bi elements. The starting materials were put into a small alumina crucible and sealed in a quartz tube in Argon gas atmosphere. The tube was then heated to 1050 °C for 2 days, followed by a subsequently cooling down to 400 °C at a rate of 3 °C/h. The plate-like single crystals as large as a few millimeters can be obtained. The composition and structure of these single crystals were checked using Energy-dispersive X-ray spectroscopy and X-ray diffraction measurements.

### Magnetotransport and Hall effect Measurements

The magnetoresistence measurements were performed with a four-probe method. The low field measurements are performed using a 9T Physics Property Measurement

System (PPMS, Quantum Design). The high field measurements were conducted in the 31 T resistive magnet and the 45T T hybrid magnet at National High Magnetic Field Laboratory (NHMFL) in Tallahassee.

The in-plane Hall resistance $R_{zx}$ (Fig. 3a) was also measured using a four probe method in PPMS. Due to slightly asymmetric electric contacts, a small but finite interlayer resistance $R_{zz}$ is also involved in the measured $R_{zx}$. At low fields where the magnetoresistance is very small, $R_{zz}$ is nearly angular-independent and acts only as a constant background which leads to a shift in $R_{zx}(\theta)$, as shown in Fig. 3b. At high fields when the magnetoresistance becomes significant, its strong angular dependence greatly affects the measurements for $R_{zx}(\theta)$. Fortunately, $R_{zz}$ scales only with the strength of the in-plane field component $|B_{xy}|$ and $R_{zz}(\theta) = R_{zz}(360°-\theta)$, while the in-plane Hall resistance follows $R_{zx}(\theta) = -R_{zx}(360°-\theta)$, $R_{zx}$ can be separated from $R_{zz}$ by symmetrizing the data: $R_{zx}(\theta)=[R_{zx}(\theta) - R_{zx}(360°-\theta)])/2$, as presented in Fig. 3c.

**Neutron Scattering Measurements**

Single crystal neutron diffraction measurements were performed on HB-3A four-circle diffractometer with the neutron wavelength $\lambda = 1.005$ Å at High Flux Isotope Reactor at Oak Ridge National Laboratory, and the data were refined with the FULLPROF [31].

**Determination of Berry Phase for Dirac cones away from the $E_F$**

In YbMnBi$_2$, the Dirac cones with the nodes located away from the FL [2] lead to the observed SdH oscillations with two major fundamental frequencies (Fig. 1f). For such multi-frequencies oscillations, Berry phase cannot be obtained from the commonly used LL fan diagram, but is accessible through the direct fit of the oscillation pattern to the multiband LK formula [25], in which the observed SdH oscillations are treated as the linear superposition of several single-frequency oscillations. Each single-frequency oscillations can be described by the Lifshitz-Kosevich formula which takes Berry phase into account for a Dirac system [23,24]:

$$\frac{\Delta\rho}{\rho(B=0)} = \frac{5}{2}\left(\frac{B}{2F}\right)^{1/2} \frac{2\pi^2 k_B T m^*/\hbar eB}{\sinh(2\pi^2 k_B T m^*/\hbar eB)} e^{2\pi^2 k_B T_D m^*/\hbar eB} \cos[2\pi(\frac{F}{B}+\gamma-\delta)]$$

where $T_D$ is Dingle temperature, $\gamma = \frac{1}{2} - \frac{\phi_B}{2\pi}$ and $\phi_B$ is Berry phase, $\delta = 0$ and $\pm 1/8$ for the 2D and 3D systems respectively. In our fit, the oscillation frequencies $F$ and the effective masses $m^*$ for each band are taken as known parameters, obtained from the analyses in Fig. 1d. The higher harmonic components are not considered in out fits for simplicity.

**Acknowledgement**

The work at Tulane is supported by the U.S. Department of Energy under EPSCoR Grant No. DE-SC0012432 with additional support from the Louisiana Board of Regents. The work at the National High Magnetic Field Laboratory, is supported by the NSF grant No. DMR-1206267, the NSF Cooperative Agreement No. DMR-1157490 and the State of Florida. X. K. acknowledges the start-up funds from Michigan State University. The work at ORNL was sponsored by the Scientific User Facilities Division, Office of Science, Basic Energy Sciences, U.S. Department of Energy.

**Author Contributions**

The single crystals used in this study were synthesized and characterized by J.L. The magnetotransport and Hall measurements in PPMS were carried out by J.L., S.R. and L.S. The high field measurements at NHMFL were conducted by J.L., D.G., Y.S., S.C., and C.L.. J.H. analyzed the data and wrote the manuscript with the input from J.L. and Z.M. T.Z., M.Z., X. K. and H.C. performed neutron scattering experiments and data analyses. All authors read and commented on the manuscript. The project was supervised by J.H. and Z.M.

**Figure captions**

**Figure 1 | Structural, magnetic, and in-plane magnetotransport properties of YbMnBi$_2$. a,** Crystal and magnetic structure of YbMnBi$_2$. **b,** Temperature dependence of the ordered moment, measured by neutron scattering experiments. **c,** Temperature dependence of the in-plane ($\rho_{xx}$) and out-of-plane ($\rho_{zz}$) resistivity. **d,** Normalized magnetoresistance MR= [$\rho_{xx}(T,B)$-$\rho_{xx}(T,B=0)$]/$\rho_{xx}(T, B=0)$ as a function of magnetic field. The field is applied along the out-of-plane direction (*c*-axis) SdH oscillations are visible above 15T at low temperatures. Inset: MR at higher temperatures up to 125K. **e,** The fits of the FFT amplitudes of the oscillatory component of $\rho_{xx}$ to the temperature damping factor of the LK formula, which yields effective mass of 0.24m$_0$ for both frequencies. The inset shows the FFT spectrum. **f,** Oscillatory component of $\rho_{xx}$, obtained by subtracting the background, as a function of the inverse of magnetic field 1/*B* at *T*=2K and 18K. The solid lines show the fits to the two-band LK model.

**Figure 2 | Interlayer transport properties due to zeroth LL for YbMnBi$_2$. a,** Schematic for the interlayer tunneling of relativistic fermions from the zeroth LL. **b,** The field dependence of the out-of-plane resistivity, $\rho_{zz}(B)$, under different field orientations at *T*=2K. The inset shows the experimental setup. The solid lines superimposed on the data represent the fits to the tunneling model (see text). The fit for $\theta$=90° is not available since the zeroth LL disappears for in-plane field. **c,** Angular dependence of magnetoresistance, measured under different fields up to 31T and at *T*=2K. The blue lines superimposed on the data collected at *B* = 3-31T represent the fits to the model which

assumes the coexistence of the relativistic fermion tunneling and momentum relaxation channels (see text), while the red solid lines superimposed on the data collected at B = 0.1-1 T represent the fits to the $\sin^2\theta$ dependence expected for the classical orbital effect. The inset shows the fit to the $B$=31T data using the momentum relaxation model alone (see text), which is less satisfactory than the fit to the model which considers both tunneling and momentum relaxation channels.

**Figure 3 | Interlayer Hall effect for YbMnBi$_2$. a,** The experimental setup for the Hall effect. The in-plane (*x*-axis) Hall voltage is measured with applying out-of-plane (*z*-axis) current, with the magnetic field rotating on the *zy*-plane. **b,** Angular dependence of the Hall resistance for $B$=0.5T. At such a low field, the LL quantization is not significant, $R_{zx}$ is proportional to the transverse field component $B\sin\theta$, as shown by the solid curve. **c,** Dependence of the in-plane Hall resistance $R_{zx}$ on the field orientation angle $\theta$. At higher fields, $R_{zx}$ is clearly deviated from the $B\sin\theta$ dependence. **b,** $R_{zx}$ plotted against $\tan\theta$, which is found to follow the same $\tan\theta$ asymptote (*i.e.* the dashed line) at low angles.

Figure 1

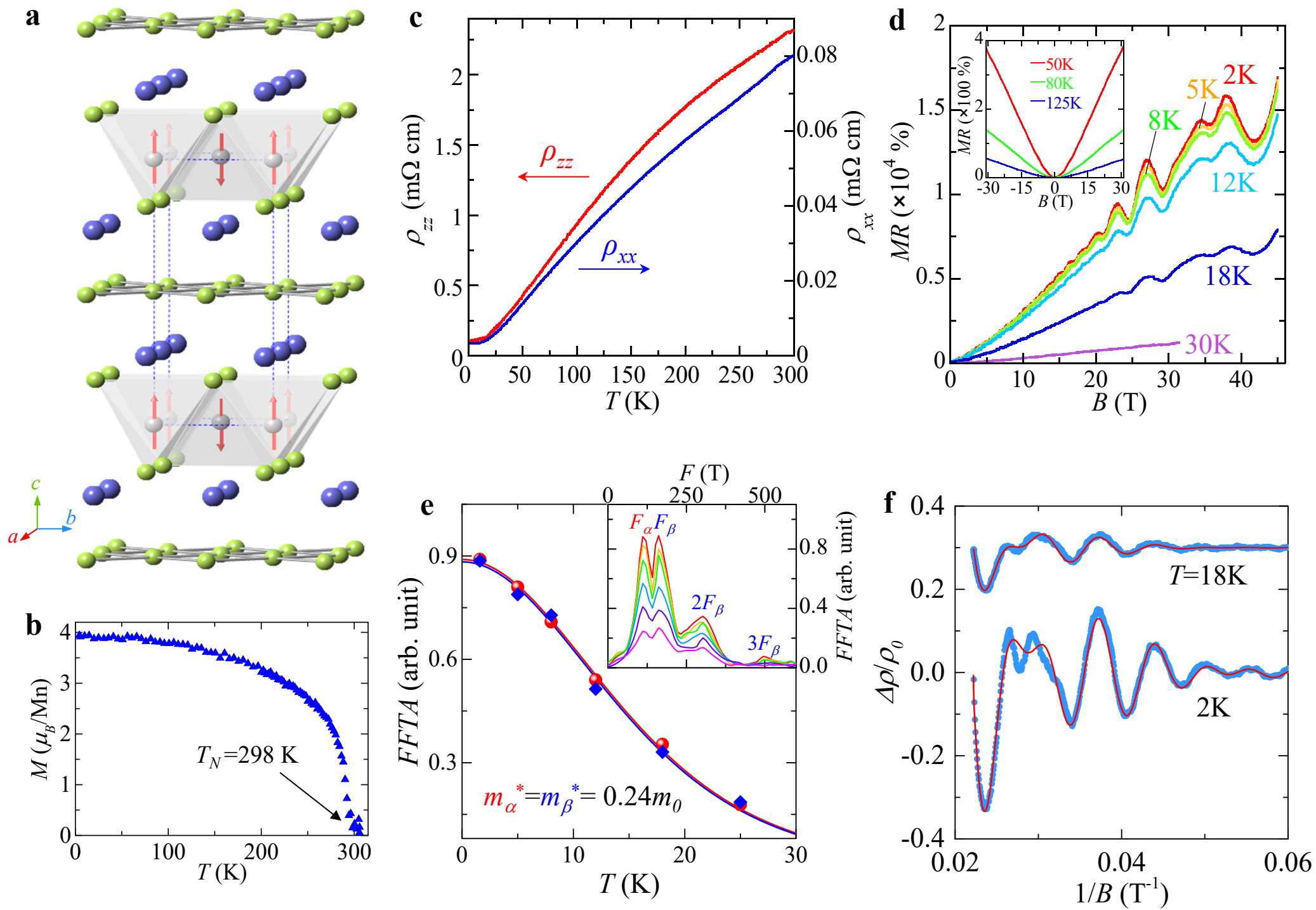

Figure 2

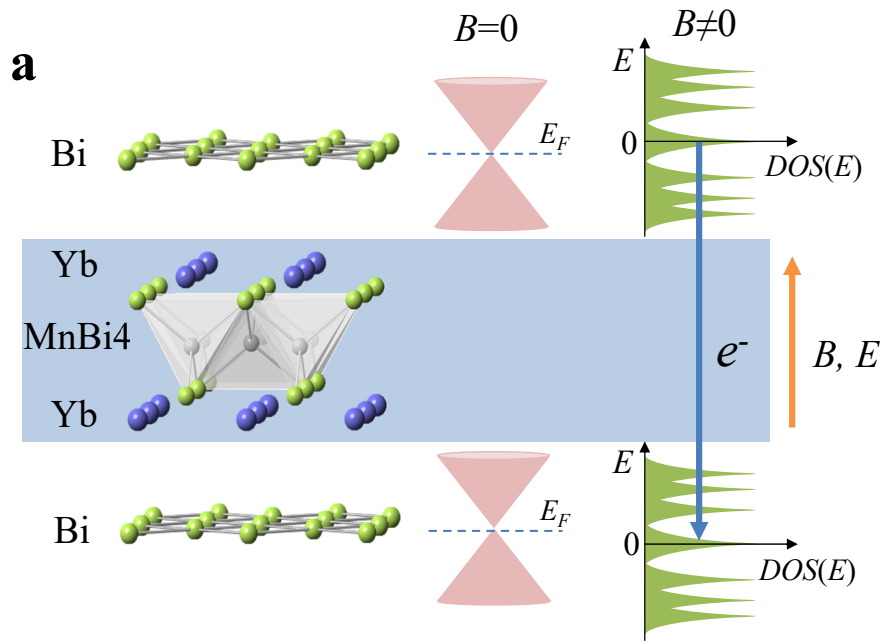
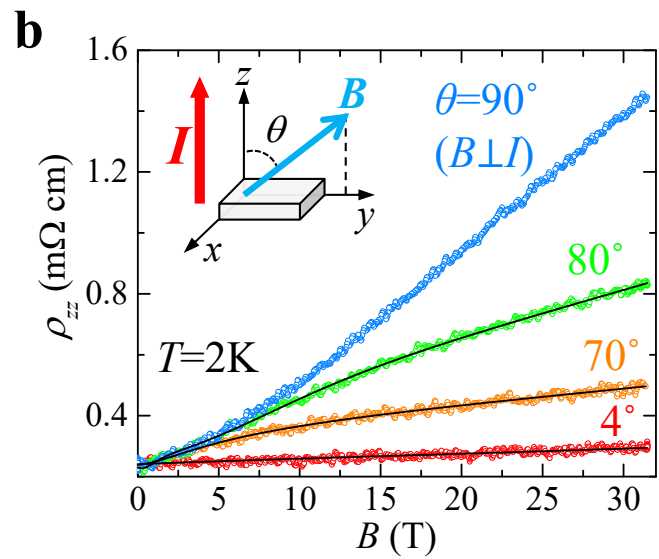
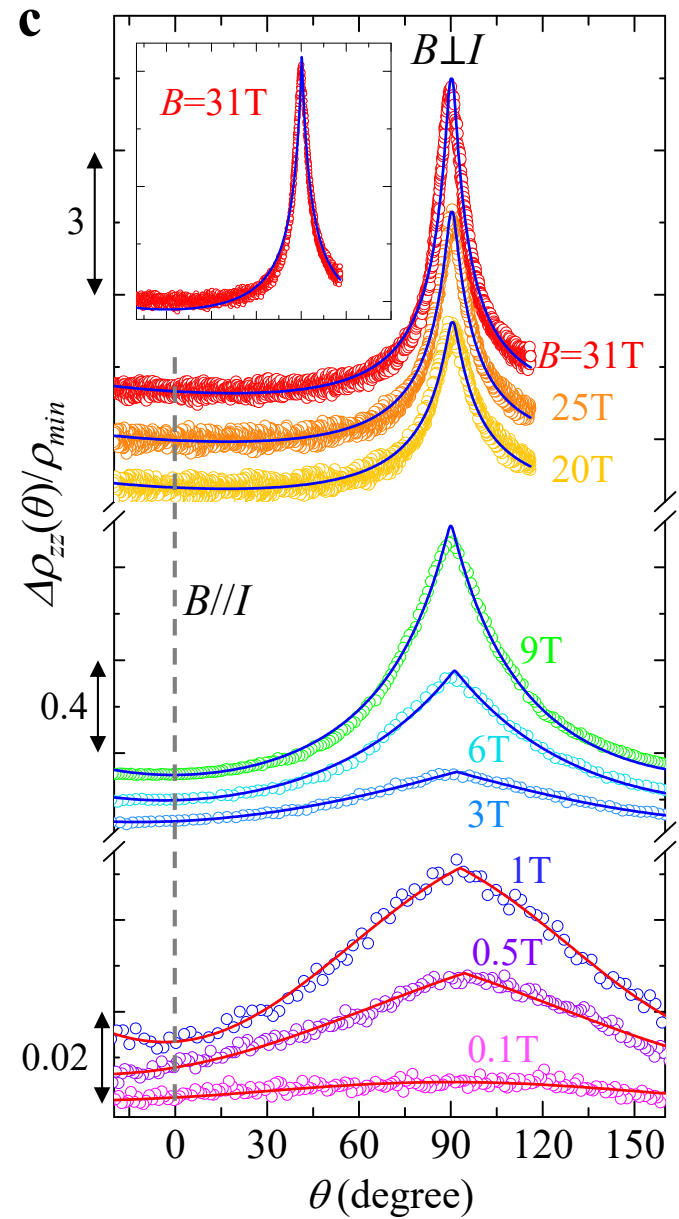

# Figure 3

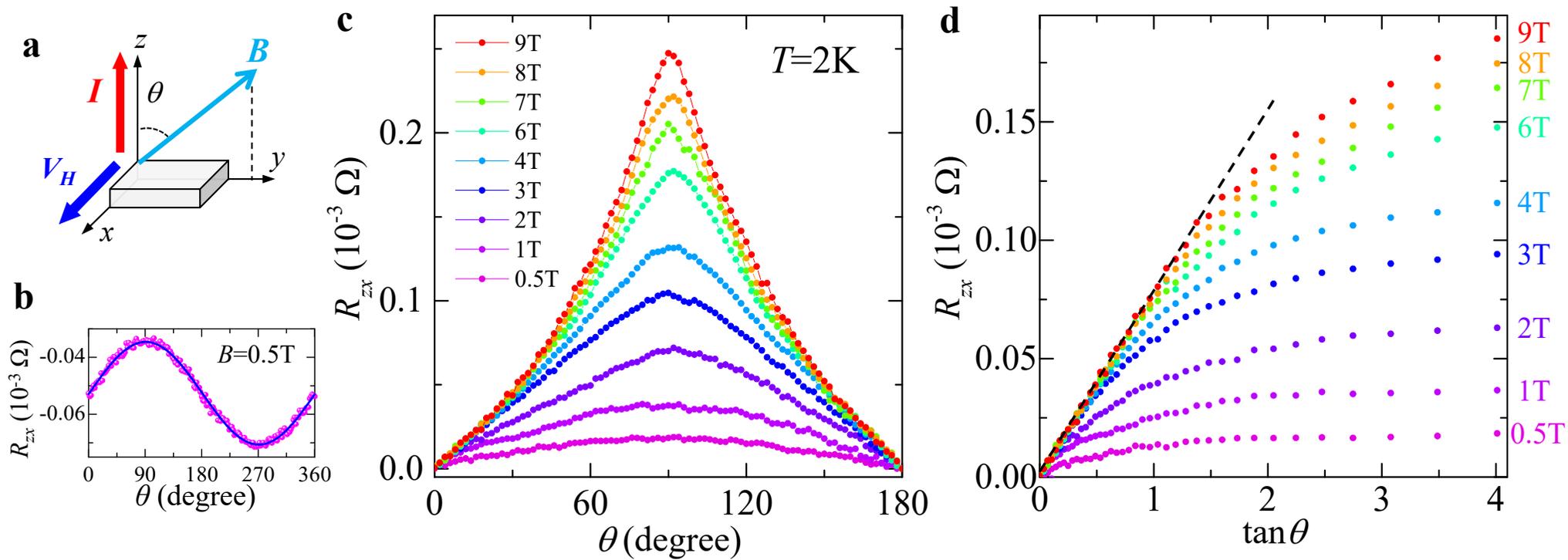

Supplementary Information

**Supplementary Table 1** | Lattice parameters of YbMnBi$_2$ at $T = 4$ K, obtained from single crystal neutron scattering measurements with the goodness of fit $\chi^2 = 0.130$.

| | | \multicolumn{5}{c}{Space Group $P4/nmm$, $a = b = 4.460$ Å, $c = 10.824$ Å} | | | |
|---|---|---|---|---|---|---|
| | | $x$ | $y$ | $z$ | Occupancy | $B_{iso}$ |
| Atom coordinates | Yb | 0.000 | 0.500 | 0.73173(56) | 1 | 0.16351(135) |
| | Mn | 0.000 | 0.000 | 0.000 | 1 | 0.23617(338) |
| | Bi1 | 0.000 | 0.000 | 0.000 | 1 | 0.08873(150) |
| | Bi2 | 0.000 | 0.500 | 0.16592(82) | 1 | 0.17722(151) |

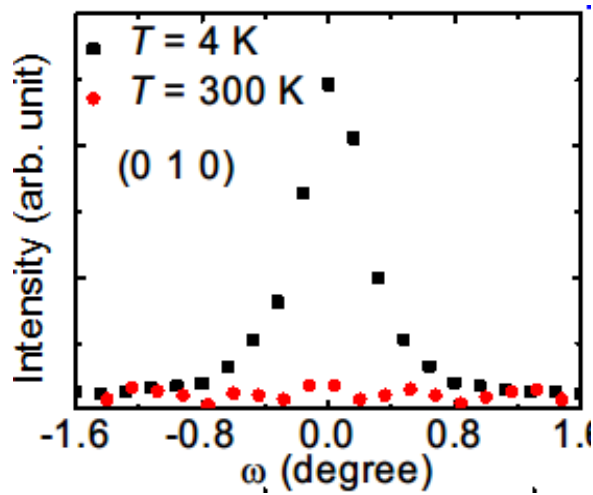

**Supplementary Figure 1 | Rocking curve scan of the (0 1 0) Bragg peak.** A Gaussian-shape peak is clearly observed at $T = 4$ K but not at $T = 300$ K, which is a characteristic of magnetic diffraction. Representational analysis using the BasIrreps program in FULLPROF [1] suggests that the Mn spin structure depicted in Fig. 1a is symmetry compatible and the best fit to the data as indicated by the small $\chi^2 = 0.130$ (for refining both nuclear and magnetic structure simultaneously).

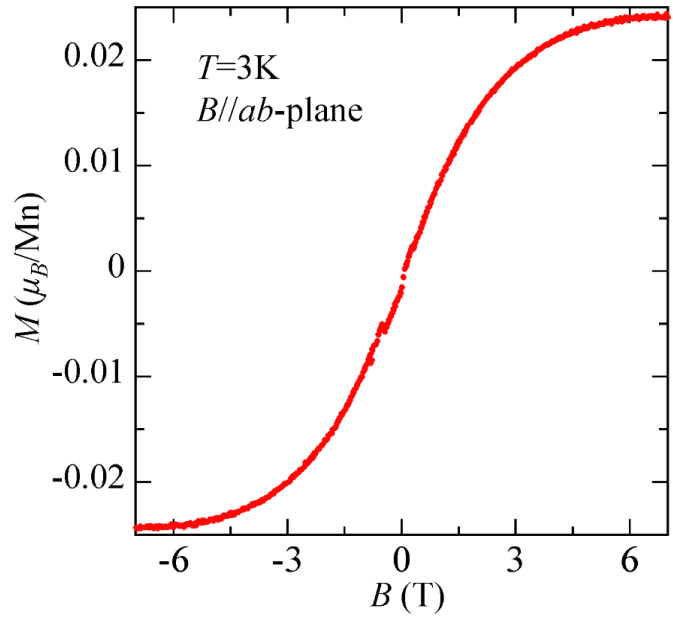

**Supplementary Figure 2 | Ferromagnetic behavior seen in the isothermal magnetization measurement of YbMnBi$_2$.**

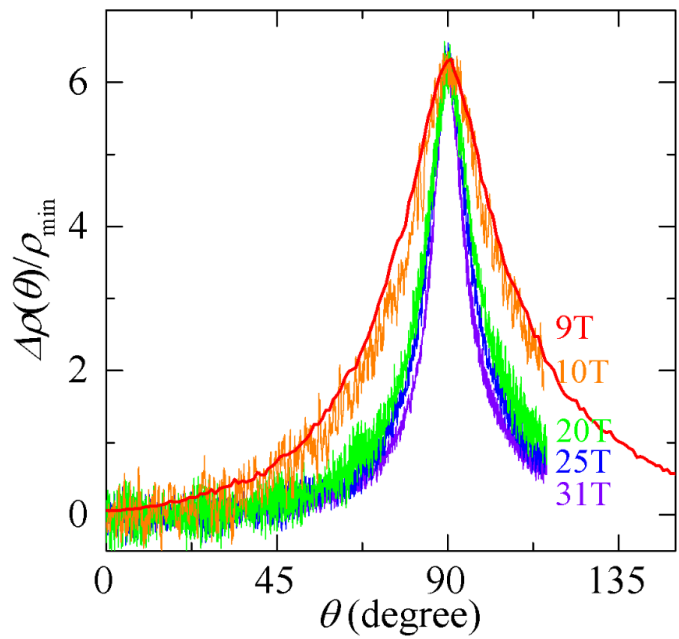

**Supplementary Figure 3 | Peak width of the AMR.** To better illustrate the evolution of peak width with field, the AMR data $\Delta\rho(\theta)/\rho_{min}$ taken at 9T, 10T, 20T, and 25T are normalized to the

peak of the $B$=31T data, *i.e.* the low-field data are multiplied by a factor such that its peak matches with that of the 31T data. The peak width is clearly dependent on the magnetic field, inconsistent with the scenario of the coherent peak [1,2].

**Supplementary Note 1 | Exclusion of other possible mechanisms for the unusual AMR anisotropy observed in YbMnBi$_2$**

Although there are several known mechanisms which can result in unusual anisotropy in MR, none of them can offer reasonable interpretation for our observed MR anisotropy shown in Fig. 2c. We first considered the spin scattering mechanism since YbMnBi$_2$ shows an AFM order near room temperature. However, the spin scattering in AFM states generally results in a sine square dependence in AMR [3], clearly inconsistent with our observation in the high field range. Second, we have examined the MR anisotropy caused by the quantum interference effects, *i.e.* weak localization (WL) and weak antilocalization (WAL). For WAL, a sharp dip in MR near $\theta$=0° is expected in our experiment setup due to quick suppression of quantum interference by the transverse field [4], which is not seen in our AMR data shown in Fig. 2c. In the case of WL, though a sharp resistance peak around $\theta$=90° is expected [5], our observed positive MR (Fig. 2b) is contradictory to the negative MR expected for WL. Another possible mechanism we have considered is the "coherent peak" originating from the formation of small closed [1] or self-crossing orbits [2] on the side of the corrugated Fermi surface under in-plane field ($B_{xy}$), as seen in the Dirac semimetal SrMnBi$_2$ [6]. However, this geometric effect should lead to a resistivity peak with a field-independent width, inconsistent with our observation of the broadening of the peak with increasing the field (see Supplementary Fig. 3). Moreover, the evolution of MR($\theta$) from a sharp peak near $\theta$ = 90° to a sine-square dependence with decreasing magnetic field does not fit to any of the mechanisms discussed above.